\documentstyle[12pt]{article}
\def\baselinestretch{1.2}
\catcode`\@=11

\@addtoreset{equation}{section}
\def\ksection{\arabic{section}}

\def\@normalsize{\@setsize\normalsize{15pt}\xiipt\@xiipt
\abovedisplayskip 14pt plus3pt minus3pt%
\belowdisplayskip \abovedisplayskip
\abovedisplayshortskip  \z@ plus3pt%
\belowdisplayshortskip  7pt plus3.5pt minus0pt}

\def\small{\@setsize\small{13.6pt}\xipt\@xipt
\abovedisplayskip 16pt plus3pt minus3pt%
\belowdisplayskip \abovedisplayskip
\abovedisplayshortskip  \z@ plus3pt%
\belowdisplayshortskip  7pt plus3.5pt minus0pt
\def\@listi{\parsep 4.5pt plus 2pt minus 1pt
            \itemsep \parsep
            \topsep 9pt plus 3pt minus 3pt}}

\def\underline#1{\relax\ifmmode\@@underline#1\else
	$\@@underline{\hbox{#1}}$\relax\fi}
\@twosidetrue

\catcode`@=12

\catcode`\@=11

\def\thesection{\Roman{section}.}

\def\FERMIPUB{}
\def\FERMILABPub#1{\def\FERMIPUB{#1}}
\def\ps@headings{\def\@oddfoot{}\def\@evenfoot{}
\def\@oddhead{\hbox{}\hfill
	\makebox[.5\textwidth]{\raggedright\ignorespaces --\thepage{}--
	\hfill {\rm FERMILAB--Pub--\FERMIPUB}}}
\def\@evenhead{\@oddhead}
\def\subsectionmark##1{\markboth{##1}{}}
}

\ps@headings

\catcode`\@=12

\relax

\newcounter{appendix}

\def\appendix{\par
 \addtocounter{appendix}{1}
 \def\thesection{Appendix \Alph{appendix}:}
 \def\ksection{\Alph{appendix}}}

\newskip\humongous \humongous=0pt plus 1000pt minus 1000pt

\newif\ifdtup

\def\oldreffmt#1{\rlap{[#1]} \hbox to 2\parindent{}}

\def\figfmt#1{\rlap{Figure {#1}} \hbox to 1in{}}

\def\bra#1{\left\langle #1\right|}
\def\ket#1{\left| #1\right\rangle}
\def\VEV#1{\left\langle #1\right\rangle}

\def\beq{\begin{equation}}
\def\eeq{\end{equation}}
\def\bea{\begin{eqnarray}}
\def\eea{\end{eqnarray}}
\def\half{\frac{1}{2}}

\def\bq{\begin{quote}}
\def\eq{\end{quote}}

\def\half{\frac{1}{2}}
\def \lta {\mathrel{\vcenter
     {\hbox{$<$}\nointerlineskip\hbox{$\sim$}}}}
\def \gta {\mathrel{\vcenter
     {\hbox{$>$}\nointerlineskip\hbox{$\sim$}}}}

\relax
\begin{document}
\par \vskip .05in
\FERMILABPub{93/397--T}
\begin{titlepage}
\begin{flushright}
FERMILAB--PUB--93/397--T\\
December, 1993\\
Submitted to {\em Phys. Rev.} {\bf D}
\end{flushright}
\vfill
\begin{center}
{\large \bf Top Production: \\
Sensitivity to New Physics }
 \end{center}
  \par \vskip .1in \noindent
\begin{center}
{\bf Christopher T. Hill} and
{\bf Stephen J. Parke }
  \par \vskip .02in \noindent
{Fermi National Accelerator Laboratory\\
P.O. Box 500, Batavia, Illinois, 60510
\footnote{ Electronic addresses: (internet)
hill@fnal.fnal.gov, parke@fnal.fnal.gov.
}}
  \par \vskip .02in \noindent
\end{center}
\begin{center}{\large Abstract}\end{center}
\par \vskip .01in
\begin{quote}
The production cross--section and distributions
of the top quark are sensitive to
new physics, e.g., the $t\overline{t}$ system can be
a probe of new resonances or gauge bosons that are strongly
coupled to the top quark, in analogy to Drell--Yan production.
The existence of such new physics is expected in
dynamical electroweak symmetry breaking schemes,
and associated with the large mass of the top quark.
The total top production cross--section can be
more than doubled, and distributions significantly
distorted with a  chosen scale
of new physics of $\sim 1 $ TeV in the vector color
singlet or octet $s$--channel.
New resonance physics is most readily discernible in the high--$p_T$
distributions of the single top quark and of the $W$ boson.
\end{quote}
 \par \vskip .02in \noindent

\vfill
\end{titlepage}
\def\baselinestretch{1.6}
\tiny
\normalsize

\noindent
{\bf I. Introduction}
\vskip .1in

\noindent
The top quark has proven to be very much heavier than most
people anticipated over a decade ago, and is now
the only expected sequential fermion with a
mass of order the electroweak symmetry breaking scale.
Since all fermion Dirac masses in the standard model must
necessarily arise from the very mechanism that breaks electroweak
symmetry, the top quark is the entity, whose existence is
assured, that is
most strongly coupled to the symmetry breaking dynamics.
The top quark may therefore prove to be a powerful probe of
electroweak symmetry breaking physics.

Through the top quark we can imagine three ways in which new
physics begins to make an appearance \cite{H1}:

\noindent
\begin{quote}
(i) oblique radiative corrections, as in
the comparison of $m_t$ to $M_W$ or
$\sin^2\theta$, etc., may yield inconsistencies in
high statistics, precision tests;

\noindent
 (ii) the search for exotic decay modes, such as
$t\rightarrow H^+ + b$, may yield the first observation
of scalars as in multiple Higgs doublet schemes or some
of the plethora of SUSY particles;

\noindent
(iii) the production distributions
of top are potentially sensitive to exotic intermediate heavy
states, such as new resonances or gauge bosons, which may favor
strong coupling  to the top quark.
\end{quote}

\vskip .1in
\noindent
In this paper we focus on the latter issue, since it will
be easiest to place some interesting new
limits immediately upon the discovery of
top, rather than through the other programatic modes
which require high statistics.  The essential idea here is
to view the $t\overline{t}$ system as a replacement for the
$\mu\overline{\mu}$ system in Drell--Yan--like processes, with
the motivation that heavy top may act as a special probe of
new associated electroweak symmetry breaking physics.

Several candidate new physics contributions that might
significantly modify top production come to
mind. These include particles of extended technicolor \cite{TC}
such as
color singlet scalars and color nonsinglet resonances,
(into which we could include generalized Higgs bosons),
new color singlet vector mesons
(which include new $U(1)$'s, new $SU(2)$'s, and ``techni-$\rho$'s'')
,
possible large induced effects in walking extended technicolor
\cite{ETC}, such
as strongly coupled extended massive
techni--bosons, and  scalar SUSY partners.
Many of these are expected to
be reasonably strongly coupled to the top quark.  For example, walking
ETC requires that some special strong dynamics in the ETC
sector be provided to generate
the top quark mass \cite{ETC}. Top quark condensation
is the extreme limit in which new dynamics generates the top quark
mass,  simultaneously breaking the electroweak
symmetry \cite{BHL}, which may also imply the
existence of new strong dynamics at the TeV scale \cite{H}.

 At $\sqrt{s}=1.8$ TeV, top quark production is dominated by
quark--anti-quark annihilation in a $p\overline{p}$ machine
\cite{QCD1}.
The actual cross--section increases over the next--to--leading
order approximation by about $30\%$ when soft--gluon
radiation effects are summed \cite{QCD2}.  Thus, we can only say with
reasonable confidence
that the overall cross--section is known to about $\pm 50\%$ in QCD.
The kinematic distributions should be reliable, however, up to their
overall normalization since they involve very large $p_T$, and
are known to be reliably determined in other processes.

Our present strategy
is two--fold. We first
take a general approach, e.g., as in ref.\cite{ELP},
writing down a complete set of
allowed $d=5$ and $d=6$ operator contact terms, subject
to the relevant symmetries, and include their
effects into top quark production.  This is the usual approach one
takes in considering new physics possibilities
from a bottom--up perspective, and we find it to be instructive.
The operator basis turns out to be severely restricted by electroweak
and custodial chiral
symmetry,
$SU(2)_L\times SU(2)_R$, in the light sector.
We show that large effects can occur at the Tevatron for
contact terms
which are additive, incoherent color singlet $s$--channel
operators, or which add
coherently to the  $q\overline{q}$ annihilation
amplitude, involving a color--octet $s$--channel. The
contact terms yield approximately only an overall renormalization of total
cross-section, and the kinematic distributions are not significantly
distorted. A signal of distortion requires the nontrivial propagator
poles of new resonances.

We then discuss the actual dynamics of hypothetical new resonances.
Motivated by our contact term analysis, we consider color singlet and
color--octet vector states.
We  first describe these resonances generally, and
then specialize to particular models, e.g.,
one in  which new octet vector resonances are massive gauge bosons.
With resonances we see that dramatic distortions
of the kinematic production distributions
can occur, and we identify the top quark $p_T$ distribution and the
decay $W$ boson $p_T$ distribution as sensitive observables.
These distributions can distinguish between the various
cases, e.g., color singlet vs. color octet, etc.
We will not provide an
encyclopedic study of all possibile masses
and couplings of new resonances, but rather we will illustrate
some generic cases which we believe are
most representative.  The key point that production studies of top
may reveal new signals, and that limits will
immediately emerge with the discovery of top,
will be illustrated.  Remarkably, we find
that we will already be sensitive
to the $\sim 1$ TeV scale with $\sim 100$ top pairs in hand.

\vskip 0.1in
\noindent
{\bf II. Contact Terms }
\vskip .1in

The problem of classifying the allowed $d\leq 6$ contact terms is
equivalent to the problem of specifying the relevant symmetries that apply.
We are interested in operators that mediate transitions between the
light quarks $(u,d)$, as well as the gluon $G_{\mu\nu}^a$, and
the heavy top quark, or third generation. We are also treating the
electroweak theory in the broken phase and ignoring weak transitions.
As such, it is appropriate to assume
that the following symmetries are linearly realized:
(i) all local gauge symmetries $SU(2)_L\times U(1)\times SU(3)$
apply linearly to the
light quarks; (ii) custodial $SU(2)_R$ applies to the light
fermions; (iii) we treat the top quark as a sterile singlet under
$SU(2)_L\times SU(2)_R$; (iv) electric  $U(1)$ invariance applies
to all fields.
Essentially, (i) and (ii)
is a massless or chiral limit for all of
the light fermions in the electroweak
theory, i.e., this would be the exact global symmetry
of the standard model if the light quark and leptons have vanishing
masses.  Conditions (iii) and (iv) reflect the fact that we
are interested in the broken phase of the electroweak theory
(but not electrodynamics) and the
top quark has a mass of order the weak scale.

 Thus, we organize the light quarks into $SU(2)_L\times
SU(2)_R$ doublets as $\psi_L=(u,d)_L$ and $\psi_R=(u,d)_R$.
The $d=6$ four--fermion operators may be written down upon
considering the possible quantum numbers
of new hypothetical force carriers, and  we indicate
the quantum numbers of the
force carriers in $(SU(2),\; SU(3))$ notation.
The symmetry assumptions powerfully restrict the
$d=6$ four--fermion operators
to the following:

\vskip 0.1in
\noindent
Singlet $s$--channel, $(1,1)$:
\bea
{\cal{O}}^1_{L,R} & = &
(\overline{\psi}_{L,R}\gamma_\mu \psi_{L,R})
( \overline{t}\gamma^\mu t); \qquad
{\cal{O}}^2_{L,R}  = (\overline{\psi}_{L,R}
\gamma_\mu \psi_{L,R})( \overline{t}\gamma^\mu\gamma^5  t )
\nonumber \\
\eea
Flavor singlet, color--octet, $s$--channel,
$(1,8)$:
\bea
{\cal{O}}^3_{L,R} & = & (\overline{\psi}_{L,R}\gamma_\mu
\frac{\lambda^a}{2} \psi_{L,R})( \overline{t}\gamma^\mu
\frac{\lambda^a}{2} t );\qquad
{\cal{O}}^{4}_{L,R}  =  (\overline{\psi}_{L,R}\gamma_\mu
\frac{\lambda^a}{2}\psi_{L,R})( \overline{t}\gamma^\mu
\gamma^5 \frac{\lambda^a}{2} t)
\nonumber \\
\eea
Here the flavor indices and fermion color indices are contracted
within the parentheses, unless otherwise specifically noted.
 Note that the $SU(2)_L\times SU(2)_R$
linear chiral symmetry precludes all but current--current interactions
(e.g., an operator such as $(\overline{\psi}_{L}
\psi_{R})( \overline{t}t)$
cannot be singlet under $SU(2)_L\times SU(2)_R$).
This
is really a statement of custodial $SU(2)$ protectionism,
e.g., an operator
of the form $\overline{q} q \overline{t} t$ would threaten to induce
a large mass for $q$ via
the induced $\bra{0}\overline{t} t\ket{0}\sim m_t v^2$,
where $v$ is the electroweak scale, and is thus disallowed
by $SU(2)_L\times SU(2)_R$.  Such operators could occur in
reality, but would be
suppressed by a power of $m_q$ which we take to be small.

There are many other $d=6$ operators we can write down which
are consistent with the symmetries, but are all reducible
by Fierz rearrangement back into
the basis ${\cal{O}}^1_{L,R}$ through ${\cal{O}}^4_{L,R}$.
The additional operators are as follows:
\vskip 0.1in
\noindent
$s$--channel,
$(\half ,1)$; $s$--channel,
$(\half , {\bf 8} )$:
\bea
{\cal{O}}^5_{L,R} & = & (\overline{\psi}^i_{L,R}\Gamma_{(x)}^A  t)
( \overline{t}\Gamma^A{}^{(x)} \psi_{i,\;L,R} );
\qquad
{\cal{O}}^6_{L,R}  =  (\overline{\psi}^i_{L,R}\Gamma_{(x)}^A
\frac{\lambda^a}{2} t)
( \overline{t}\frac{\lambda^a}{2}\Gamma^A{}^{(x)}
\psi_{i,\;L,R} )\qquad
\nonumber \\
\eea
 The $\Gamma_{(x)}^A$ are generic
Dirac matrices where $(x)$ represents the summed Lorentz
indices and
$A$ is an unsummed  type label designating
scalar ($A\equiv S$), vector ($A\equiv V$), tensor (
$A\equiv T$), axial--vector ($A\equiv A$), and
pseudoscalar ($A\equiv P$).

We furthermore note that $\psi$ may be replaced
by its electroweak and Dirac
charge--conjugate $\psi^C= i\tau^2 C \overline{\psi}^T$,
where $C= i\gamma^0\gamma^2$.
This replacement must be done pairwise
since we want to conserve $U(1)_{EM}$, and we note that the
following operators can also be reduced back to
the original basis ${\cal{O}}^1_{L,R}$ through
${\cal{O}}^4_{L,R}$:
\vskip 0.1in
\noindent
 $t$--channel,
$(\half , {\bf \overline{3}})$; $t$--channel,
$(\half , {\bf {6}})$:
\bea
{\cal{O}}^7_{L,R} & = & (\overline{\psi}^{Ci}_{L,R\;[\alpha }
\Gamma_{(x)}^A  t_{\beta ]})( \overline{t}^{[\alpha}
\Gamma^A{}^{(x)}  \psi^{C \;\beta ]}_{i,\;L,R} );
 \;\;\;\;
{\cal{O}}^8_{L,R}  =  (\overline{\psi}^{Ci}_{L,R\;\{\alpha }
\Gamma^A_{(x)}  t_{\beta \}})( \overline{t}^{\{\alpha}
\Gamma^A{}^{(x)}  \psi^{C \;\beta \}}_{i,\;L,R} )\qquad
\nonumber \\
\eea
where $\alpha$ and $\beta$ are color indices,
appropriately symmetrized
or antisymmetrized.

To see that ${\cal{O}}^5_{L,R}$ through ${\cal{O}}^8_{L,R}$
are reducible to ${\cal{O}}^1_{L,R}$ through ${\cal{O}}^4_{L,R}$,
we first observe that operators like
${\cal{O}}^{}_{L,R}  =  (\overline{\psi}^i_{L,R}
\Gamma_A t)( \overline{t}\gamma^5  \Gamma^A\psi_{i,\; L,R}) $
are equivalent to
the ${\cal{O}}^{5}_{L,R}$ set upon commuting $\gamma^5$ through $\Gamma^A$,
and using the chiral projections (similarly for the color--octet analogues).
Also, we need only
consider as independent the $S$, $V$, and $T$ combinations, since
$A$ and $P$ are equivalent by the same argument.
Moreover, the operators are not independent under Fierz rearrangement.
For example, the $LL$ tensor operators vanish by chirality
(similarly for $RR$):
\bea
(\overline{\psi}^{i\alpha}_{L}\sigma_{\mu\nu}t_\alpha)
(\overline{t}^\beta\sigma^{\mu\nu}\psi_{i,\beta; L})
& = &  -3(\overline{\psi}^{i\alpha}_{L}\psi_{i,\beta; L})
(\overline{t}^\beta
t_\alpha)  -
\half(\overline{\psi}^{i\alpha}_{L}\sigma_{\mu\nu} \psi_{i,\beta; L})
(\overline{t}^\beta \sigma^{\mu\nu} t_\alpha)
 \nonumber \\
& & \left.
- 3(\overline{\psi}^{i\alpha}_{L}\psi_{i,\beta; L})
(\overline{t}^\beta \gamma^5 t_\alpha) \right)
 \nonumber \\
& = & 0
\eea
The remaining
$S$ and $V$ can be related to ${\cal{O}}^1_{L,R}$
through ${\cal{O}}^4_{L,R}$, e.g., as in
\beq
(\overline{\psi}^i_{L,R}t)(\overline{t}\psi^i_{L,R})
= -\frac{1}{2} \left( \overline{\psi}_{L,R}\gamma_\mu
\frac{\lambda^a}{2} \psi_{L,R} \overline{t}\gamma^\mu
\frac{\lambda^a}{2} t  \pm
\overline{\psi}_{L,R}\gamma_\mu \frac{\lambda^a}{2}
\psi_{L,R} \overline{t}\gamma^\mu\gamma^5 \frac{\lambda^a}{2} t
+ {\cal{O}}(1/N_c)
\right).
\eeq
with the $+$ ($-$)  sign for the $LL$ ($RR$) case
(keeping terms beyond leading order in $1/N_c$
does not affect the argument).

In the case of the operators  ${\cal{O}}^7_{L,R}$ and
 ${\cal{O}}^8_{L,R}$, the color of
$\psi^C$ is conjugated, which leads to color
$t$--channels $\bf \overline{3}$ or $\bf 6$. However,
the same Fierz reduction arguments apply to this set as before.
We can always treat $\psi$ as a generic two--component electroweak
doublet, Dirac field and at the end of the rearrangement
replace it by $\psi^C$.
Thus, in the case of $O^7$, we will  obtain by Fierz
rearrangement:
\bea
O^7 & \rightarrow & (\overline{\psi}^{C}_{ [\alpha }
\Gamma^A \psi^{C\; \beta]})
( \overline{t}^{[\alpha } \Gamma^A  t_{\beta]} )
\;\;\;
 \rightarrow  \;\;\;
 (\overline{\psi}^{C}\gamma_\mu \left(\frac{\lambda^{A}}{2}
\right)^T\psi^{C\; })
( \overline{t}\gamma^\mu \frac{\lambda^{A}}{2} t ) \; + \; ...
\nonumber \\
& \rightarrow &  (\overline{\psi}\gamma_\mu \frac{\lambda^{A}}{2}
\psi^{\; })
( \overline{t}\gamma^\mu \frac{\lambda^{A}}{2} t ) \; + \; ...
\eea
and this set is also equivalent
to ${\cal{O}}^1_{ L,R}$ through ${\cal{O}}^4_{ L,R}$.
Physically, in $q\overline{q}$ annihilation we have only
the possibility of  $s$--channel singlet and octet;
$t$--channel ${\bf \overline{3}}$ and ${\bf {6}}$ exchange
is indistinguishable from $s$--channel singlet and octet
exchange at low energies.
Thus, the combination of light quark electroweak and custodial
symmetries together with the top
treated as an effective sterile singlet yields a
large reduction of the operator basis that we would have to consider otherwise.

There are, in addition, the $d\leq 6$ gluonic operators,
\bea
{\cal{O}}^1_{g}  =   {m_t}G_{\mu\nu}^a \overline{t}
\sigma^{\mu\nu}\lambda^a  t; \qquad &&
\qquad
{\cal{O}}^2_{g}  =  {m_t}\tilde{G}_{\mu\nu}^a \overline{t}
\sigma^{\mu\nu}\lambda^a t; \nonumber \\
{\cal{O}}^3_{g}  =   (D^\mu G_{\mu\nu})^a \overline{t}\gamma^\nu\lambda^a  t;
\qquad && \qquad
 {\cal{O}}^4_{g}  =
 (D^\mu G_{\mu\nu})^a \overline{t}\gamma^\nu
\gamma^5 \lambda^a t;
\eea
${\cal{O}}^{3}_{g}$ and ${\cal{O}}^{4}_{g}$ are penguin--like and
equivalent by use of equations of motion
to the four--fermion operator set. $d=5$ operators of this form
might be expected from
strong--ETC radiative corrections to the top--top--gluon vertex.
We therefore include a factor of $m_t$ in their definition,
which is the natural scale of these operators (with an arbitrary
$\frac{1}{\Lambda}$ coefficient in the effective
Lagrangian we would end up with
unnaturally large contributions of the effects of these terms;
if they are large in production dynamics, then they would also be expected
to produce a large induced $m_t$ as well).
At the Tevatron we find that there is little
sensitivity to the gluonic contact
terms from the subprocess $ g + g \rightarrow \overline{t} + t$
in the top quark production rate.
The gluonic operators can contribute to the top quark
production in the subprocess  $\overline{q} + q
\rightarrow \overline{t} + t$.
However the extra factors of $m_t$ suppress these
contributions compared with the previous four fermion operators.
Previous authors have considered the effect of the color--octet
pseudo--Nambu--Goldstone boson \cite{P,G} and anomalous gluonic magnetic
moment operators (as described below) in $\overline{t}t$ production.
However, these effects were considered only in the gluon
fusion subprocess, i.e., the leading process
in a $pp$ machine such as SSC or LHC.  In general these gluon fusion
processes are a small effect for us.
 We shall not
consider them further presently, and
refer the reader to ref.\cite{P} and ref.\cite{G}.

The color singlet $s$--channel operators give incoherent
additive corrections
to the QCD annihilation cross--section. On the other hand,
 color--octet $s$--channel operators must be added coherently to the
single gluon annihilation amplitude in
$\overline{q}
+ q \rightarrow \overline{t} + t$.
Let us consider the particular effective interaction Lagrangians:
\beq
{\cal{L}}_1' = \frac{g_3^2}{\Lambda^2}\left( {\cal{O}}^1_L
+  {\cal{O}}^1_R \right)
\qquad\qquad
{\cal{L}}_2' = \frac{g_3^2}{\Lambda^2}\left( {\cal{O}}^3_L
+  {\cal{O}}^3_R \right)
\eeq
where we include a factor of the QCD coupling constant, $g_3^2$,
which will be seen to be convenient when we compare to actual
resonance models
as described in the next section ($g_3^2$ is a smaller normalization
factor than used in
ref.\cite{ELP}, who assume a
strong coupling constant $g^2\sim 4\pi$;
we take typically $g_3^2 \sim 4\pi\times  (0.11)$; therefore our scale
$\Lambda$ will always be $\sim 1/3$ of the scale obtained by using the
ref.$\cite{ELP}$ normalization).
We have modified the standard tree level top production calculation
\cite{KS} to include the contact terms in ${\cal{L}}_i' $ and have
used the HMRS Set-B parton
distribution functions \cite{HMRS} with both the factorization
and renormalization scales set equal to $m_t/2$.

We see in Figures 1(a) and 1(b) the results for the ratio
of the total top production cross--section,
to the QCD cross--section,  with various scales $\Lambda$ (thus $R=1$ is
$\Lambda\rightarrow \infty$, or pure QCD).  We have plotted
these as functions of $m_t$. The effects of the singlet operators
are shown in Fig. 1(a), while octets are shown in Fig. 1(b).
 With the signs chosen for the color octet contact
terms of ${\cal{L}}_2'$  we generate coherent additive enhancement of top
production (denoted in Fig. 1(b) by $+$); with the opposite sign (denoted $-$)
for the color--octet $s$--channel operators, i.e., ${\cal{L}}_2' \rightarrow
-{\cal{L}}_2'$ we suppress the production cross--section.
Clearly, the enhancement  or suppression
of the cross--section by the contact term effect
increases with increasing $m_t$ for fixed $\Lambda$ because the gluon
propagator is decreasing with increasing $m_t$ while the contact term
remains constant.

In Fig.(2) we plot the differential distributions $d\sigma/d p_T$ for
the $p_T$ of the single top quark system
and the $p_T$ of the $W$ boson produced by a decaying top quark.  The $W$
boson  $p_T$ is readily known since the lepton and
missing $p_T$ will be known in the $\ell + 4 j$ sample.
 In the  $\ell + 4 j$ sample it should also be possible to cut
the $3j$ invariant mass near the top quark mass and determine
the $3j$ $p_T$ to infer the
distribution of the top quark with $p_T$.
We have given these distributions because we have
found that they are empirically the most sensitive
to new resonance physics, as described
below, thus optimizing the potential observability of new
physics.    However, we see that
when the new physics is given only by a contact term, for fixed $m_t$,
the QCD distributions are only approximately renormalized by an
overall multiplier,
i.e., no significant distortion of the distributions is seen.
Thus, these distributions are shaped
primarily by the phase space and kinematics.  This is a problem for the
observability of mere contact term effects, since one must trust the
knowledge of the overall QCD production cross--section normalization.
Nonetheless, if we discover that top is produced with a total cross-section
that is significantly larger or smaller than the QCD predictions of
refs.\cite{QCD1,QCD2},
then it is possible that one is seeing the effects of new physics.  For
example, if the scale of the contact term is $\Lambda \leq 600$ GeV
then the total top cross--section is increased
relative to pure QCD by a factor greater than $2$.

\vskip 0.1in
\noindent
{\bf III. A New Strong Dynamics }
\vskip .1in
The contact term analysis reveals a possible
sizeable enhancement of top production through new color--singlet or
color--octet $s$--channel vector mesons.
This may involve either
a new intermediate force carrier in the
$s$--channel, or $t$--channel, ${\bf \overline{3}}$
or ${\bf 6}$ exchange, by Fierz rearrangement.
The contact terms lead only to
multiplicative renormalization of the kinematic distributions,
and are therefore not easy to distinguish, except by their
very large effects when $\Lambda\lta 0.6$ TeV.  We turn now
to dynamical  resonances which are generally expected
to have influence in
kinematic distributions.

A new color singlet $s$--channel resonance $B_\mu$ of mass $M_B$,
can be either
a boundstate, as in technicolor models,  or a new gauge boson,
such as a heavy $Z'$ gauge field.  For example, the model of
Lindner and Ross \cite{H} attempts to generate top condensation
with such a field, and favors strong coupling to top.
We will presently confine our attention to vector current
couplings, in which case the coupling to the
light quark doublets and the top quark can be taken as:
\beq
g_3(\tilde{z}_1\overline{\psi}\gamma_\mu\psi
+ \tilde{z}_2\overline{t}\gamma_\mu t) B^{\mu}
\eeq
where $\tilde{z}_1$ and $\tilde{z}_2$ are just
 scale factors for the interaction strength
relative to QCD (we use the $\tilde{}$ for
the color singlet case). The width of the
singlet vector boson to decay to  $n_f$ flavors of light quark
pairs,  and to $t\overline{t}$ in the massless approximation (for the
$M_B$ values we consider the massless approximation
is valid even for $m_t\sim 160$ GeV) is:
\bea
\Gamma_B = \frac{g_3^2 M_B}{12 \pi}(n_f\tilde{z}_1^2 + \tilde{z}_2^2)
\eea
In general, the presence of the
vector boson will modify the QCD production
cross--section for the
parton subprocess $\overline{q} + q \rightarrow \overline{t} + t$
as:
\beq
\sigma_{\overline{q} + q \rightarrow \overline{t} + t}
= \sigma_{QCD}\left[ 1 + C_3\frac{\tilde{z^2_1}\;\tilde{z^2_2} \;s^2}{
[(s-M_B^2)^2 + M_B^2\Gamma_B^2]}  \right]
\eeq
where $C_3$ is an $SU(N)$ color factor given by $4N^2/(N^2-1) \sim
4.5$.

We also consider a color octet of massive vector bosons, $B_\mu^A$,
also of mass $M_B$, which
we will call ``colorons.''    These are necessarily degenerate because
QCD is unbroken.  Such objects may be composite, $\rho$--like objects,
with a scale of compositeness of order $\sim M_B$
which serves as a cut--off scale for the $B_\mu^A$ effective
Lagrangian. It then suffices to consider $B_\mu^A$ as a linear
representation of QCD (if $B_\mu^A$ are fundamental, with effectively
infinite cut--off, then consistency requires they are
necessarily gauge bosons, as we consider below).
We can write a phenomenological
coupling as:
\beq
g_3(z_1\overline{\psi}\gamma_\mu\frac{\lambda^A}{2}\psi
+ z_2\overline{t}\gamma_\mu\frac{\lambda^A}{2}t) B^{A,\mu}
\eeq
Here we again use rescaled QCD couplings, $z_1g_3$
and $z_2g_3$.
The process $\overline{q} + q \rightarrow \overline{t} + t$ involves
the coherent sum of
the gluon and coloron $s$--channel amplitudes.  It is therefore easy
to implement the effect of the coloron in QCD production, i.e.,
we simply make the replacement in the gluon propagator:
\beq
\frac{g_3^2}{s}\;\; \rightarrow\;\; \frac{g_3^2}{s}
 + \frac{g_3^2 z_1z_2}{(s - M_B^2) + iM_B\Gamma_B}
\eeq
The production amplitude therefore depends upon the
product of the new rescaled coupling constants, $z_1z_2$
(which can be negative), the mass $M_B$, and
the coloron width $\Gamma_B$. As we will see below, the relative sign
of these terms, determined by the sign of $z_1z_2$, has a significant
effect in shaping the top quark $p_T$ distributions.

The width of the coloron to decay to  $n_f$ flavors of light quark
pairs,  and to $t\overline{t}$ in the massless approximation (for the
$M_B$ values we consider the massless approximation
is valid even for $m_t\sim 160$ GeV) is:
\bea
\Gamma_B = \frac{g_3^2 M_B}{24 \pi}(n_fz_1^2 + z_2^2)
\eea
Obviously, the width can be readily generalized.
Thus, in the generic phenomenological model,
top production depends upon $3$ independent parameters,
$z_1z_2$, $M_B$ and $\Gamma_B/M_B \sim (n_fz_1^2 + z_2^2)$.

It is useful to reduce the parameters
by focusing on a particular version of the general model which is
dictated by simple additional symmetry assumptions.  Thus, we turn to
the case in which $B$ is a gauge boson.
Let us consider a simple scheme in which the third generation
participates in a new strong gauge dynamics, analogous to QCD.  This has been
previously suggested by
the fact that the top quark is very heavy \cite{H}, implying
either strong coupling
to the electroweak breaking dynamics, or  associated with
that breaking through ``top condensation.''
Our present model is a slightly simpler variation on ref.\cite{H}.

We assume a minimal extension of the standard model
such that at scales $\mu \gg M$, we have the gauge group
$U(1)\times SU(2)_L \times SU(3)_1\times SU(3)_2$. The coupling constants
(gauge fields) of $SU(3)_1\times SU(3)_2$ are now respectively
$h_1$ and $h_2$ ($A^A_{1\mu}$ and $A^A_{2\mu}$).
We assign the usual light quark and lepton fields to representations under
$(SU(2)_L, SU(3)_1, SU(3)_2)$ as follows:
\bea
(u,\; d)_L,\;\; (c,\; s)_L & \rightarrow & \; (2,3,1) \qquad\qquad
u_R, \; d_R, \; c_R, \; s_R,  \rightarrow  \; (1,3,1)
\nonumber \\
(\nu_e,\; e )_L,\;\; (\nu_\mu,\;\mu )_L, \;\; (\nu_\tau,\; \tau )_L
 & \rightarrow  & \; (2,1,1)
 \qquad\qquad
e_R, \; \mu_R, \; \tau_R, \;( \nu_{iR} )  \rightarrow  \; (1,1,1)
\nonumber \\
\eea
while the top and bottom quarks are assigned:
\beq
(t,\; b)_L  \rightarrow  \; (2,1,3);\qquad
t_R, \; b_R  \rightarrow \; (1,1,3)
\eeq
This assignment is anomaly free.

We break the symmetry $SU(3)_1\times SU(3)_2
\rightarrow SU(3)_c$ at the scale $M$.   We can do so by introducing
a scalar (Higgs) field $\Phi^{a}_{b'}$
which transforms as $(1,3,\bar{3})$.
By a general choice of the $\Phi$ potential
a VEV develops: $\VEV{\Phi} = \makebox{diag}(M,\; M,\; M)$.
breaking  $SU(3)_1\times SU(3)_2$ to the massless
QCD gauge group $SU(3)_c$ with gluons, $A_\mu^A$
and a residual global $SU(3)'$ with degenerate, massive
colorons, $B_\mu^A$.
The gluon $(A_\mu^A)$
and coloron $(B_\mu^A)$ fields are then defined by:
\beq
A^A_{1\mu}  = \cos\theta A^A_\mu - \sin\theta B^A_\mu \qquad
A^A_{2\mu}  = \sin\theta A^A_\mu + \cos\theta B^A_\mu
\eeq
where:
\beq
h_1\cos\theta = g_3;\qquad  h_2\sin\theta = g_3;\qquad
\eeq
and thus:
\beq
\tan\theta = h_1/h_2;\qquad \frac{1}{g_3^2} = \frac{1}{h_1^2} +
 \frac{1}{h_2^2}
\eeq
where $g_3$ is the QCD coupling constant at $M$.
In what follows we envision $h_2 \gg h_1$ and thus $\cot\theta \gg 1$
(perhaps to select the top quark direction for condensation).
The mass of the degenerate octet of colorons is given by:
\beq
M_B = \left(\sqrt{h_1^2 + h_2^2}\right) M
= \left(\frac{g_3}{\sin \theta \cos \theta}\right) M
\eeq

The usual QCD gluonic interactions
are obtained for all quarks (including top and
bottom)
while the coloron interaction takes the form:
\bea
{\cal{L}}' & =  & -\left[ g_3\cot\theta
\left( \bar{t}\gamma_\mu \frac{\lambda^A}{2} t +
 \bar{b}\gamma_\mu \frac{\lambda^A}{2} b \right)
- \; g_3\tan\theta \sum_i\bar{q}_i\gamma_\mu \frac{\lambda^A}{2} q_i
\right] B^{\mu A}
\eea
where the sum extends over all other light quarks.
The appearance of the $\cot\theta$ factor in the light quark
sector and the $-\tan\theta$ in the heavy quark sector implies
that the production amplitude for top via annihilation
$\overline{q} + q \rightarrow \overline{t} + t$ is independent
of $\theta$, except through the
coloron width.  Note that we have now reduced the number of
free parameters from $3$ to $2$, i.e., $\cot\theta$ and $M_B$,
since now $z_1z_2=-1$.
Again, top production involves the coherent sum of
the gluon and coloron $s$--channel amplitudes:
\beq
\frac{g_3^2}{s}\rightarrow \frac{g_3^2}{s}
- \frac{g_3^2}{(s - M_B^2) + iM_B\Gamma_B}
\eeq
Now a relative minus sign between
the coloron and gluon contributions
is fixed implying that the cross--section  is
amplified for $s < M_B^2$ (neglecting the $B$ width),
and suppressed
for $s > M_B^2$.   This model involves only the free parameters
$\cot\theta$ and $M_B$, and the sensitivity to  $\cot\theta$
appears only through the width of the coloron $B$.

The decay width of the coloron is dominated by $t\overline{t}$
for large $\cot\theta$:
\bea
\Gamma_B = \frac{g_3^2\cot^2\theta}{24 \pi}
M_B
\eea
If top-condensation occurs, or if the coloron plays a role in
inducing a large top quark mass through near critical coupling,
then $\cot\theta$ is roughly determined.
Criticality implies $g_3\cot\theta = h >> 1$;
for example, the Nambu--Jona-Lasinio value of
$h$ sufficient to trigger condensation is
$h^2N_c/8\pi \gta 1$. This yields  $h^2 = g^2_3\cot^2\theta \gta 8\pi/3$,
and applying this to the coloron decay width yields:
\bea
\Gamma_B \gta \frac{2}{9}M_H.
\eea
(where the factor of $2$ includes both $b$ and $t$ contributions).
We use this latter result to define the width of both
colorons and color--singlets in the
following, taking $\Gamma_B  = 0.2 \; M_B$, and
thus fixing all parameters except $M_B$.

We consider subsequently the following  three cases:
{\bf (A):} the color--singlet vector resonance model
as defined by eq.(10) in which
$z_1z_2=\pm 1$ and where we take the width $\Gamma_B=0.2\; M_B$
for definiteness; {\bf (B):} the gauge color--octet vector resonance model
as defined by eq.(22) in which
$z_1z_2=-1$; {\bf (C):} a hybrid model in which the parameters
of the gauge color--octet vector resonance model
are assumed, but we invert by
hand the relative signs between the
$\cot\theta$ and $\tan\theta$ terms of eq.(22), i.e, $z_1z_2= +1$.

In the sequence of
figures, 4(a), (b)
and (c), we present the differential distributions for the single top
quark $p_T$ in the three models. Fig. 4(a) represents the incoherent
additive contribution of the color
singlet vector resonance to the cross--section, i.e., model {\bf (A)}.
Notice the large additive coherent enhancement of model {\bf (B)}
for smaller $p_T$, as seen in Fig. 4(b), where the amplitudes
add below the mass of the resonance, while in 4(c) we give the
results for the hybrid model {\bf (C)} in which we
have reversed the relative signs of the couplings of the vector boson
to light quarks and the top quark.  Here we see an actual
coherent suppression
of the distributions for low $p_T$, followed by a characteristic sizeable
enhancement at high--$p_T$.

In Fig.(5) we present the differential distributions for the $p_T$
of the $W$ boson coming from the weak decay
of a single top.
In either the $W$ or $t$ distributions of Figs.(4) through
(5) one sees
a significant signal by the
distortion relative to QCD for resonances as heavy as
$800$ GeV.  As the resonance mass $M_B$ is taken larger than $800$ GeV
the distortions rapidly go away, due to the finite energy reach of
the machine, assumed to be $\sqrt{s}=1.8$ TeV.  Even with a few events in
hand one may see, in principle, a large excess of high--$p_T$ top
quarks if such resonances are present.

Finally, since we have argued the similarities of
this approach with Drell--Yan,
in Fig.(6) we present the differential distributions for the mass
of the
$t\overline{t}$ pair, $M_{tt}$.  Again, a significant signal of
distortion relative to QCD for resonances as heavy as
$700$ GeV is seen.

Our results have assumed essentially a process in which the
product of resonance production and decay coupling contants
is given by $g_3^2$, or $|z_1z_2|=1$ for
our three models.  This is in general a conservative guess, though
it is  forced in the special case of
model {\bf (B)}, and we emphasize that the product of
coupling constants could
be significantly larger than this in reality, i.e.,
$g_3^2$ at the scale $\sim m_t$
is really quite perturbative,
and new resonances could enter with much stronger
nonperturbative coupling constants in principle.  Even at the
level we have assumed for these couplings, we see that we are
very sensitive to the resonance profiles. In effect, when the
resonance can be excited the
pole of the propagator is becoming $\sim 1/M_B\Gamma_B$, and the effective
value of $\Lambda$ is now much smaller than $M_B$; since
we take $\Gamma_B= 0.2 M_B$ we get an effective $\Lambda \sim 0.44 M_B$
from the pole, which increases the cross--section in the
vicinity of the pole by $\sim 25$
times.  For example, in the
singlet case we can readily see the effects of a $700$ GeV resonance
in a substantial increase in the top quark $p_T$.
 This indicates that production studies of top
with large statistics would be sensitive to more weakly coupled states
of large mass than we have assumed, though the mass sensitivity drops
rapidly to zero when
$M_B\gta \half\sqrt{s}$, favoring increased collider energies.

\newpage
\noindent
{\bf IV. Conclusions}
\vskip .1in

With the discovery of the top quark we are entering a new and
previously unexplored realm of fermions that have masses
of order the weak scale, and thus couplings to
the dynamics of electroweak symmetry breaking that are $\gta O(1)$.
New physics associated with the top quark might
therefore be expected to occur, and this can first show
up in the physics of top production.

To model
this situation,  we have first considered
generic contact terms.  We think that the treatment of
the $t$ quark as a singlet and the use of $SU(2)_L\times SU(2)_R$
to constrain the light quark sector is a new approach, valid in
the broken phase of the electroweak symmetry, which leads to a dramatic
simplification of the operator basis.  We then study the effects of
such contact terms in an effective Lagrangian. We find  that large
effects can occur for the  color singlet or octet in the
$s$--channel in $q\overline{q}$ annihilation
(though physically, this can also come about from a
${\bf \overline{3}}$ or ${\bf 6}$ exchange in the $t$--channel).
A significantly enhanced or suppressed top production rate
relative to QCD is indicative of new physics of this kind,
and can occur for $\Lambda$ as large as
$\Lambda\sim 0.7$ TeV  (if we used the stronger
coupling normalization of \cite{ELP} we would be sensitive
to $\Lambda\sim 2.0$ TeV).

We are
therefore lead to
consider models of
specific strongly coupled vector resonances and
gauge bosons, which
have been previously motivated by consideration of the relationship of
the top quark to electroweak symmetry breaking.
We consider essentially three cases, {\bf (A)} a
color singlet vector
resonance; {\bf (B)} a gauge color octet vector
resonance (a slight variation on ref.\cite{H}), and
{\bf (C)} a hybrid color octet vector
resonance in which the resonant amplitude has an inverted sign.
We fix the resonance width by a strong coupling argument to be
$\Gamma_B\sim 0.2\; M_B$, and the product of
(rescaled)  couplings $|z_1z_2|= 1$ (this happens
automatically in model {\bf (B)}), leaving
only a remaining free parameter, the resonance mass, $M_B$.
We study the top and $W$ $p_T$ distributions, and the $t\overline{t}$ mass
distribution in the three models.
These three models have characteristically different signatures.
We emphasize that the results for observables that we
present are not the most general, but are reasonable first guesses
as to what might be expected if new strong dynamics is involved.
The effects we generate can be dramatic, leading to significant enhancements
of top production cross-sections for a scale of
new physics of order $1$ TeV, and to significant and
observable distortions of the $p_T$ distributions of the
$W$ boson and top system.

Qualitatively,
we see that top quarks produced with large cross--section and
excessively
large $p_T$ would be indicative of the excitation of
new resonances and may be observed with $\sim 100$  $t\overline{t}$
events in $\ell + 4$ jets.  A corollary is that the aplanarity of top events
would be expected to be smaller, since the top system is boosted to larger
$p_T$ than normally expected.  These might be the best first order indicators
of such new physics, and could conceiveably be seen with the first
handful of top events.

Certainly a new strong dynamics for electroweak symmetry breaking
in concert with the large value of $m_t$, must be taken seriously.
Meaningful limits will be readily established with fewer than $100$
top quarks in hand, and a non--null discovery potential is
evident.  With the discovery of the
top quark we may be commencing the first observation of
new physics beyond the standard model.

\newpage
\noindent
{\bf Acknowledgements}
\vskip .1in
\noindent
We thank E. Eichten, and W. Giele for useful discussions.
This work was performed at the Fermi National Accelerator Laboratory,
which is operated by Universities Research Association, Inc., under
contract DE-AC02-76CHO3000 with the U.S. Department of Energy.

\newpage
\vskip 0.1in
\noindent
{\bf Figure Captions}
\vskip .2in
\noindent
Figures 1(a), 1(b):  The ratio of total
cross--sections, $R=\sigma_{contact}/\sigma_{QCD}$,
for $t\overline{t}$ production
including the effect of contact terms,
relative to the QCD result, $R=1$. (a) is color
singlet $s$--channel ${\cal{L}}_1'$; (b) is color
octet $s$--channel ${\cal{L}}_2'$ with the coherently
additive $+$ or subtractive $-$ signs.

\vskip .2in
\noindent
Figure 2(a), 2(b)  The differential distribution, $\frac{d\sigma}{dp_T}$,
where $p_T$ is the transverse momentum of a single
top quark. $m_t=160$ GeV is assumed, and results given for
contact terms
with $\Lambda =0.6$ TeV, $\Lambda =0.8$ TeV, and
$\Lambda =\infty$ corresponding to pure QCD. (a) is color
singlet $s$--channel ${\cal{L}}_1'$; (b) is color
octet $s$--channel additive ${\cal{L}}_2'$.

\vskip .2in
\noindent
Figure 3(a), 3(b):  The differential distribution, $\frac{d\sigma}{dp_T}$,
where $p_T$ is the transverse momentum of a $W$ boson from
the weak decay of a single
top quark. $m_t=160$ GeV is assumed, and results given
as in Fig.(2).

\vskip .2in
\noindent
Figure 4(a), 4(b), 4(c):  The differential distributions,
$\frac{d\sigma}{dp_T}$,
where $p_T$ is the transverse momentum of a single
top quark, $m_t=160$ GeV, for
the following sequence: (a) the gauge color--singlet vector
resonance model {\bf (A)};  (b) the gauge color--octet vector
resonance model {\bf (B)}
(with $z_1z_2=-1$); (c) the hybrid gauge color--octet vector
resonance model {\bf (C)} (with $z_1z_2=+1$).
Results are given in each case for resonance masses
$M_B=(0.6,\; 0.7, \;0.8,\; 1.0)$ TeV, and $M_B
\rightarrow\infty$ corresponding to pure QCD.

\vskip .2in
\noindent
Figure 5(a), 5(b), 5(c):  The differential distribution,
$\frac{d\sigma}{dp_T}$,
where $p_T$ is the transverse momentum of a $W$ boson from
the weak decay of a single
top quark.  Results are presented in sequence as in Fig.(4).

\vskip .2in
\noindent
Figure 6(a), 6(b), 6(c):  The differential distribution,
$\frac{d\sigma}{dM_{tt}}$,
where $M_{tt}$ is mass of the $t\overline{t}$ quark pair.
Results are presented in sequence as in Fig.(4).

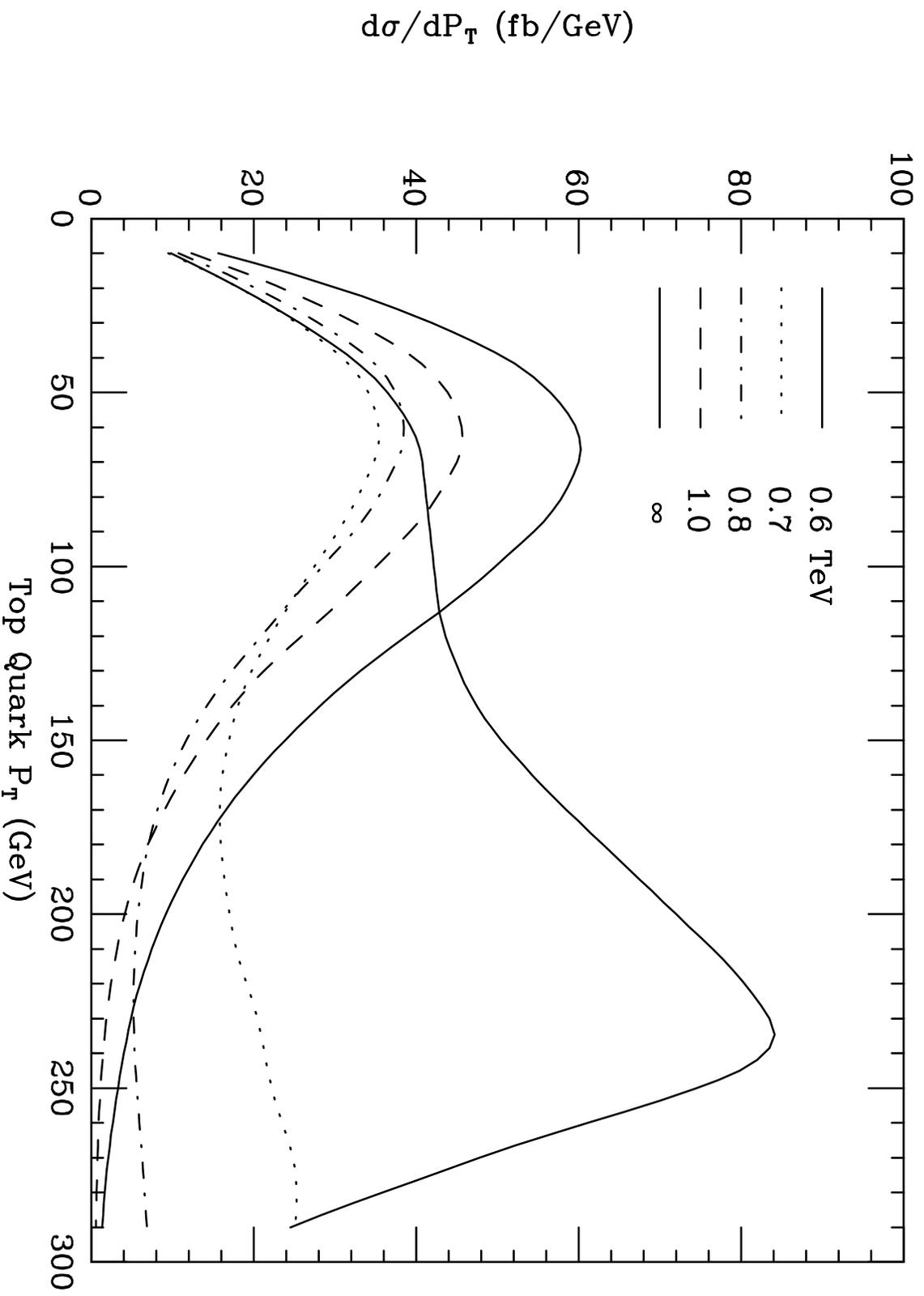
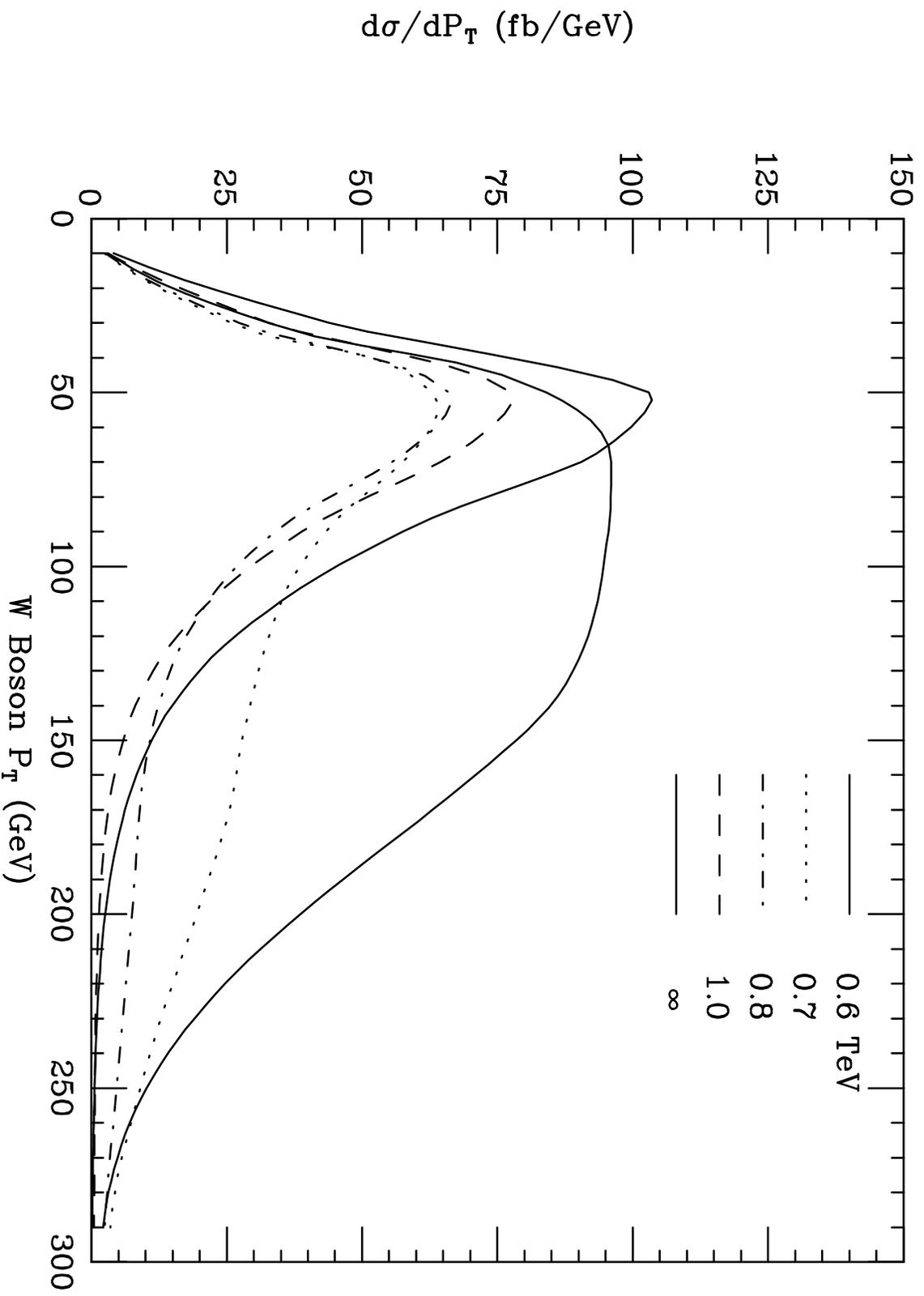
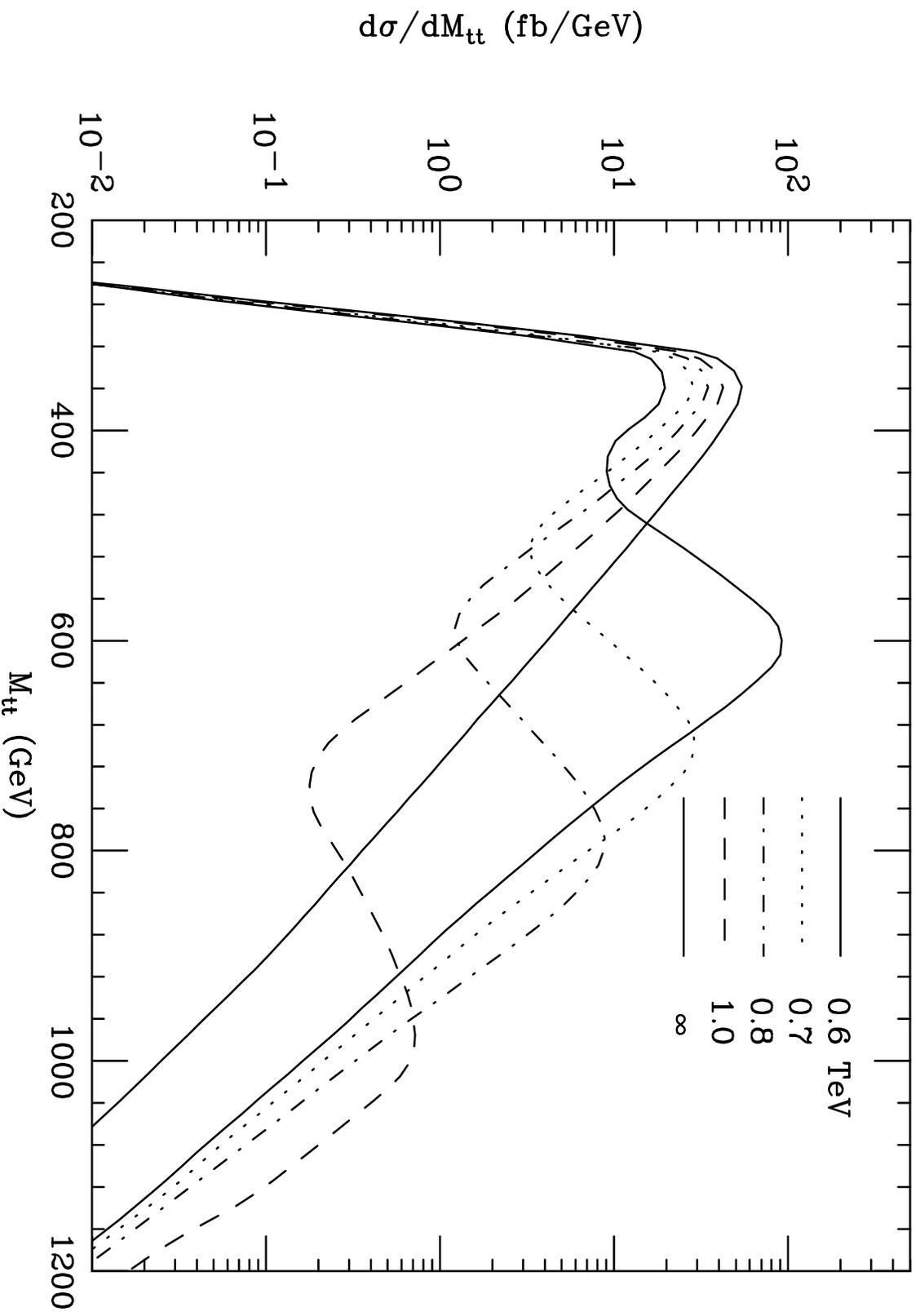
\end{document}